\documentclass[11pt,twoside]{article}
\usepackage{asp2010}

\resetcounters

\begin{document}

\title{ Comparing simultaneous measurements by two high-resolution imaging spectropolarimeters: 
the `G\"ottingen' FPI$@$VTT and CRISP$@$SST}
\author{N. Bello Gonz\'alez,$^1$ L. Bellot Rubio,$^{2,1}$ A. Ortiz,$^3$  R. Rezaei,$^1$ L. Rouppe van der Voort,$^3$ and R. Schlichenmaier$^1$ 
\affil{$^1$Kiepenheuer Institut f\"ur Sonnenphysik, Sch\"oneckstr. 6, D-79104 Freiburg, Germany}
\affil{$^2$Instituto de Astrof\'isica de Andaluc\'ia (CSIC), Apdo. 3004, 18080 Granada, Spain}
\affil{$^3$Institute of Theoretical Astrophysics, University of Oslo, P.O. Box 1029 Blindern, N-0315 Oslo, Norway}}

\begin{abstract}
In July 2009, the leading spot of the active region NOAA11024 was observed simultaneously and independently with the `G\"ottingen' FPI at VTT and CRISP at SST, i.e., at two different sites, telescopes, instruments and using different spectral lines. The data processing and data analysis have been carried out independently with different techniques. Maps of physical parameters retrieved from 2D spectro-polarimetric data observed with `G\"ottingen' FPI and CRISP show an impressive agreement. In addition, the `G\"ottingen' FPI maps also exhibit a notable resemblance with simultaneous TIP (spectrographic) observations. The consistency in the results demonstrates the excellent capabilities of these observing facilities. Besides, it confirms the solar origin of the detected signals, and the reliability of FPI-based spectro-polarimeters.
\end{abstract}

\section{Introduction}

The improvement of solar observational capabilities has revealed a wealth of unresolved structures present in the solar atmosphere whose physical driver(s) is not yet established.
For the study of these small-scale solar structures, there are various observational requirements to be fulfilled: an appropriate field of view, an angular resolution below 0.1\,arcsec, a spectral resolution of at least 300\,000 and a temporal resolution of 10\,sec or less. In view of these requirements, \cite{2007msfa.conf...67K} compares the performance of FPI- and grating-based spectrometers. The conclusion is that FPI systems are better suited when a combination of high spatial, spectral and temporal resolution is pursued.
However, the reliability of the spectro(-polarimetric) capabilities of the FPI-based instruments is still under debate, mainly due to the complexity of the optical requirements, of the processing of the large data sets acquired during the observations and other facts like the degradation of image quality intrinsic to any FPI imaging system. Yet, 2D spectro-polarimetry is a powerful technique which offers the possibility of recovering spectral line profiles in each pixel across the field of view, within a short time interval. This kind of measurements allow one to derive the spatial distribution of physical parameters such as temperature, velocity and magnetic field, together with the temporal evolution \citep{1989hsrs.conf..241B}. 
In addition, 2D filtergrams can be restored from $seeing$ effects, leading to the highest spatial resolution.

For this contribution, two independent observing teams have compared the data sets recorded simultaneously with two different FPI imaging systems in order to check the validity of their measurements, and in consequence, the performance of this kind of post-focus instruments.

\section{Observations and Data Analysis}

On 4 July 2009, two independent observing campaigns took place at the German Vacuum Tower Telescope (VTT, Observatorio del Teide, Tenerife) and the Swedish 1-m Tower Telescope (SST, Observatorio de El Roque de los Muchachos, La Palma), respectively.
The common target was the leading spot of the active region NOAA\,11024 ($\vartheta$\,$\sim$\,28$^{\circ}$). 
The observations were recorded with the `G\"ottingen' FPI \citep{2006A&A...451.1151P, 2008A&A...480..265B} at the VTT and with CRISP \citep{2006A&A...447.1111S, 2008ApJ...689L..69S} at the SST.  
The specifications of the observations presented here are collected in Table\,\ref{table}.

\begin{table}[t]
\begin{tabular}{lll}
	\hline
	\hline
 & VTT & SST   \\
  \hline  
  \hline
Location                              & Obs. del Teide, Tenerife             & ORM, La Palma\\
Telescope                          & 0.70\,m, reflector, coelostat        & 0.97\,m, refractor, turret  \\
Instrument 				       & `G\"ottingen' FPI, collimated       & CRISP, telecentric  \\
Field of view                        & $\sim$\,32"\,$\times$\,22"          & $\sim$\,55"\,$\times$\,55" \\
Spectral line   			        & Fe\,{\sc i} 617.3\,nm (g=2.5)       & Fe\,{\sc i} 630.15 \& 630.25\,nm\\
Spectral sampling 		        & 1.48\,pm, 31 points                    & 4.8\,pm, 15 +15 + 1 points \\
Exposure, accum. 		        & 20\,ms, 7 acc.                             & 17 ms, 9 acc. \\
Polarimeter 		             & full Stokes, FLCs                              & Full Stokes, NLCs \\
Image reconstruction 	        & Speckle + sp. interferometry       & MOMFBD \\
Polarimetric noise level       & 3.8\,$\times$10$^{-3}$\,I$_C$    &  1.5\,$\times$10$^{-3}$\,I$_C$ \\
	\hline
\end{tabular}
\caption{Specifications of the observations at the VTT and SST telescopes.}\label{table}
\end{table}

\subsection{Observations with the `G\"ottingen' FPI  at the VTT}

At the time of the observations, the `G\"ottingen'\,FPI\footnote{Currently updated to GFPI \citep{2011arXiv1111.5509P} and attached to the 1.5\,m GREGOR telescope.} (G-FPI, from now on) was operational at the VTT. 
The full Stokes vector was retrieved from four polarimetric measurements of the incoming sunlight taken with a modulator system, based on ferro-electric liquid crystals (FLCs), of estimated efficiencies $\epsilon_Q$=0.41,  $\epsilon_U$=0.49, and  $\epsilon_V$=0.58.
The data were recorded in speckle mode, i.e., (seven) frames of short exposure (20\,ms) per wavelength and polarimetric state. 
This procedure allows one to restore the images by applying speckle reconstruction techniques \citep{1996A&AS..120..195D} and speckle interferometry \citep{1992A&A...261..321K, 2007msfa.conf..217B}.
Further detailed information on the data and the data analysis procedure can be found in \cite{2012A&A...537A..19R}.\\

\subsection{CRISP at the SST}

CRISP, operational at the SST, 
is likewise an imaging spectropolarimeter based on two FPI etalons, yet it is a telecentric mounting, i.e., the etalons are placed close to the focal plane.
The differences between a telecentric and a collimated mounting and their optical requirements are discussed in \cite{2001AN....322..375K} and \cite{2006A&A...447.1111S}.
Together with the present data, quasi-simultaneous observations in the Fe\,{\sc i} 630.15\,nm line and other wavelengths (filters) not discussed here, were recorded.
The data were restored applying Multi-Object Multi-Frame Blind Deconvolution \citep[MOMFBD, ][]{2005SoPh..228..191V}.
CRISP measures the polarimetric states of the incoming light with a modulator system based on two nematic liquid crystals (NLCs).
The physical parameters were retrieved applying the SIR code \citep{1992ApJ...398..375R} to an area of 26$^{\prime\prime}$\,$\times$\,27$^{\prime\prime}$ centred in the spot.

\begin{figure}[t]
\begin{center}
\includegraphics[width=14cm,bb=14 14 1029 637, clip]{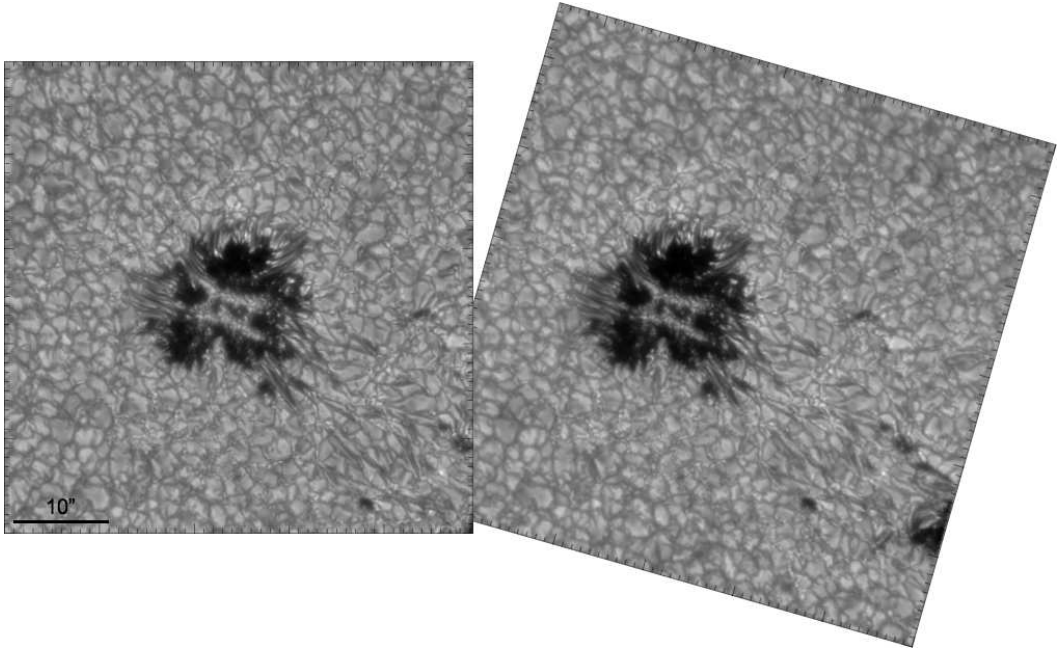}
\caption{Intensity maps of NOAA\,11024 leading spot observed at 09:50\,UT. {\em Left:} $\sim$\,50"$\times$\,50" intensity map recorded with the G-band channel at the VTT. {\em Right:} continuum (630.25\,nm) intensity map ($\sim$\,55"$\times$\,55") recorded with CRISP. For a better comparison, the CRISP image has been rotated by 74$^{\circ}$ counterclockwise in order to match the VTT image orientation.}
\label{fig:int}
\end{center}
\end{figure}

\section{Comparing the Retrieved Physical Parameters}

Let us emphasise that the observations, as the data analysis, were carried out by two independent groups with independent scientific goals, hence, the diversity and differences in the observations and data analysis methods.
The aim of this contribution is to compare qualitatively the results from the two independent approaches.

\subsection{Intensity Maps}

Simultaneously to the G-FPI scans, images in a G band channel (430 nm, 1 nm bandwidth) were recorded.
Figure\,\ref{fig:int} shows intensity maps from the G band (VTT) and narrow-band continuum at $\sim$\,630.2\,nm (SST). The quality of the images is comparable. This is expected since the diffraction limit of the VTT at 430\,nm and the SST at 630\,nm are similar. 
The higher contrast shown by the G band image (11\%) with respect to the continuum image (6\%) is inherent to the formation in the deep photosphere of the molecular band. 
Nevertheless, the filigrees and other bright fine structures characteristic of G band images can also be identified in the continuum data. 
A deeper comparison of the images after proper co-alignment  shows a one-to-one correlation, even at the smallest structures. 
This guarantees the solar origin of the observed structures and discards them to be artifacts introduced by, e.g., the image reconstruction techniques.

\subsection{Doppler Maps}

\begin{figure}[!h]
\begin{center}
\includegraphics[width=14cm,bb=14 14 1007 474]{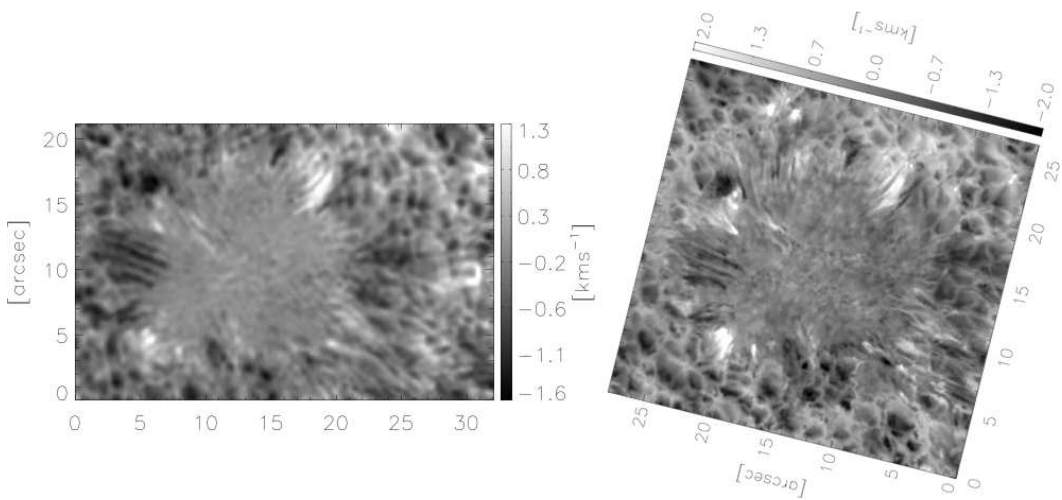}
\caption{Velocity maps of NOAA\,11024 leading spot (09:50\,UT). {\em Left:} `G\"ottingen FPI' Doppler map measured with the COG method from Stokes I profiles. {\em Right:} CRISP velocity map retrieved from inversions. The maps are clipped to the velocity values given in the color bars. For a better comparison, the CRISP map has been rotated by 74$^{\circ}$ counterclockwise in order to match the VTT map orientation.}
\label{fig:vel}
\end{center}
\end{figure}

Figure\,\ref{fig:vel} shows velocity maps from the G-FPI (left) and CRISP (right) data. 
The G-FPI map has been obtained by measuring Doppler shifts of the centre-of-gravity (COG) of the Stokes $I$ profiles (at 617.34\,nm) with respect to the average profile from the ambient granulation. The result is the line-of-sight component of the velocity field. In the case of the CRISP data, the ${v}_{LOS}$ has been obtained as an output from the inversions (at 630.25\,nm). 

Despite the difference in the spatial resolution, due to the different throughput of the instruments and the telescope apertures, and the different measurement approaches, the resemblance is evident, i.e., the correlation of the redshifts (clear areas) and blueshifts (dark areas) is consistent. Distinct features like, e.g., the redshifts observed in the G-FPI map at ($x$, $y$)\,=\,(10$^{\prime\prime}$, 12$^{\prime\prime}$) and (9$^{\prime\prime}$, 14$^{\prime\prime}$) and co-spatial with the upper light bridge of the spot, are clearly seen in the CRISP data. This confirms the reliability of the observations as well as the different data-analysis approaches.

\subsection{Line-of-sight Component of the Magnetic Field}
The LOS component of the magnetic field has been measured from the COG method \citep{1979A&A....74....1R}
applied to the $I+V$ and $I-V$ profiles for the G-FPI data. For comparison, we have calculated LOS magnetograms from the CRISP magnetic field strength and inclination maps inferred from the SIR inversions. 
Note that these maps correspond to a different moment in the sunspot formation than the previous maps. They also differ by about 2\,min between them. However, once more, the similitude of the magnetograms, clipped to the same values, leaves no doubt about the solar nature of the observed signal.

\begin{figure}[!h]
\begin{center}
\includegraphics[width=14cm,bb=14 14 1004 464]{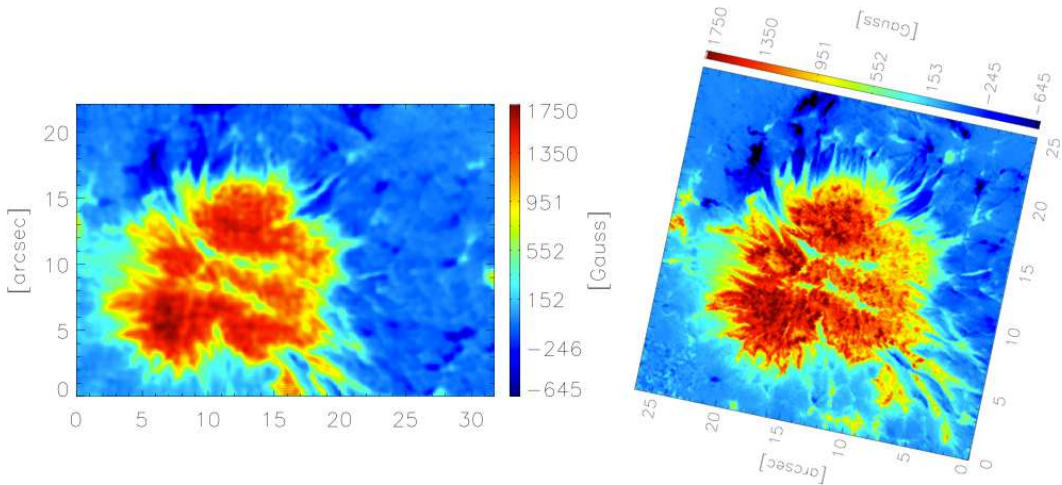}
\caption{LOS magnetograms of NOAA\,11024 leading spot. {\em Left:} `G\"ottingen FPI' LOS magnetogram measured with the COG method (10:11\,UT). {\em Right:} CRISP LOS magnetogram inferred from inversions (10:13\,UT). The maps are clipped to the values given in the color bars. For a better comparison, the CRISP map has been rotated by 74$^{\circ}$ counterclockwise in order to match the VTT map orientation.}
\end{center}
\end{figure}

\subsubsection{Comparison of 2D and 1D Spectro-polarimetric VTT Data} 
During the observations with the G-FPI, additional data with the Tenerife Infrared Polarimeter \citep[TIP II,][]{2007ASPC..368..611C}, attached to the Echelle spectrograph, were recorded in the IR wavelength range around the Fe {\sc i} line at 1089.6\,nm ($g_\textrm{\scriptsize eff}$=1.5). 
A large (spatial) scan with TIP II covered the whole sunspot while simultaneously scanning in wavelength with the G-FPI. 
The correspondence of the simultaneous and co-spatial maps of the physical parameters with the 2D and the 1D spectro-polarimeters is excellent \citep[see Figs.\,2, 3 and 4 in][]{2012A&A...537A..19R}. This is a crucial evidence of the ability of the G-FPI to measure spectro-polarimetric signals.

\section{Conclusions}

We have compared qualitatively some of the results obtained from simultaneous but independent observations which differ in many aspects (observing sites, telescopes, instruments, image reconstruction techniques and data analysis). The resemblance of the intensity, as well as the spectroscopic and polarimetric signals, despite the differences in spatial resolution intrinsic to the optical systems, is remarkable. The similitude of the different maps, even at the smallest scales, assures the reliability of the findings and validates the performance of the imaging spectro-polarimeters based on FPI etalons. 

Spectropolarimeters based on FPI imaging systems are often used today, e.g., GFPI@GREGOR,  TESOS@VTT \citep{2000A&AS..146..499V}, IBIS@DST \citep{2006SoPh..236..415C}, CRISP@SST and IMaX@Sunrise \citep{2011SoPh..268...57M}, and are planned for future large solar telescopes, e.g., VTF@ATST and the FPI systems for EST. With this contribution, we demonstrate the reliability and high-quality performance of such systems as spectro-polarimeters when the combination of highest spatial, spectral and temporal resolution is pursued.

\acknowledgements The VTT is operated by the Kiepenheuer-Institut f\"ur Sonnenphysik at the Spanish Observatorio del Teide. The SST is operated by the Institute for Solar Physics of the Royal Swedish Academy
of Sciences in the Spanish Observatorio del Roque de los Muchachos of the Instituto de Astrof\'isica de Canarias. NBG acknowledges financial support by the DFG grant Schm 1168/9-1.


\end{document}